\documentclass{article}
\usepackage{amsmath,amssymb}
\usepackage{epsfig}
\include{graphix}
\newtheorem{prop}{Proposition}
\newtheorem{defin}{Definition}
\newtheorem{rem}{Remark}
\newtheorem{eg}{Example}
\newtheorem{theorem}{Theorem}
\newtheorem{corollary}{corollary}
\begin{document}

\title{Graph Splicing System}
\author{L. Jeganathan\\ and \\
R. Rama\thanks{Corresponding author. Email :
ramar@iitm.ac.in}
\\Department of Mathematics,\\ Indian Institute of Technology Madras,\\
Chennai, India}

\date{}
\maketitle
\begin{abstract}
The string splicing was introduced by Tom Head which stands as an
abstract model for the DNA recombination under the influence of
restriction enzymes.  The complex chemical process of three
dimensional molecules in three dimensional space can be modeled
using graphs. The graph splicing systems which were studied so far,
can only be applied to a particular type of graphs which could be
interpreted as linear or circular graphs.  In this paper, we take a
different and a novel approach to splice two graphs and introduce a
splicing system for graphs that can be applied to all types of
graphs.  Splicing two graphs can be thought of as a new operation,
among the graphs, that generates many new graphs from the given two
graphs.  Taking a different line of thinking, some of the graph
theoretical results of the splicing are studied.
\end{abstract}
\section{Introduction}

To understand and analyze well the complex structure as well as the
evolutionary process of genes, researchers have long been searching
for syntactical models. One such model was a grammatical model
provided by formal language theory \cite{searls}.  Yet, the grammar
types in the Chomsky hierarchy was inadequate in describing the
biological systems \cite{collado-vides}.

\par In his pioneering work, Tom Head has proposed an operation
called `Splicing' for describing the recombinant behavior of
double-stranded DNA molecules \cite{head} which established a new
relationship between formal language theory and the study of
informational macromolecules.  Splicing operation is a formal model
of the recombinant behavior of DNA molecules under the influence of
restriction enzymes and ligases.  Informally, splicing two strings
means to cut them at points specified by the  given substrings
(corresponding to patterns recognized by restriction enzymes) and to
concatenate the obtained fragments crosswise (this corresponds to
the ligation reaction).  Since then, the theory of splicing has
become an  interesting area of formal language theory, where results
of  splicing systems on string languages (splicing systems  were
later renamed as H-systems to indicate the originator)   gave new
insights in some (closure) properties of families of string
languages \cite{paun1}. The mathematical study of the splicing
operation on the strings has been investigated exhaustively, which
lead to a language generating device viz., Extended H-systems
(EH-systems) using the `splicing operation' as the basic ingredient
\cite{paun2}. Several control mechanisms were suggested in
increasing the computing power of EH systems with finite components,
equivalent to the power of Turing machines.  Thus, splicing
operation on strings has lead to universal computing device
(programmable DNA Computers based on splicing).
\par
A splicing operation contains splicing rules of the form $(u_1,u_2 ;
u_3,u_4)$, where $u_1,u_2,u_3, u_4$ are strings over some alphabet
$V$.  We apply the splicing rule to two strings $x_1u_1u_2x_2$,
$y_1u_3u_4y_2$, ( $x_1,x_2,y_1,y_2$ are strings over $V^\star$).  As
a result, the new strings $x_1u_1u_4y_2$ and $y_1u_3u_2x_2$ are
obtained.  We  use the modified definition of splicing as it appears
in \cite{paun1}

\par DNA sequences are three dimensional objects in a
three-dimensional space.  Some problems arise when they are
described by one-dimensional strings. So, the other models of
splicing were explored.    In
\cite{culikII,rama1,rama2,krithivasan}, array splicing systems were
studied. In \cite{freund},\cite{santhanam} graph splicing systems
were discussed. But these systems cannot be applied to the graphs
that cannot be interpreted as linear or circular graphs.  Hence, we
take a different approach to splicing two graphs and introduce a
splicing system for graphs which can be applied to all graphs.
Splicing two graphs can be thought of as a new operation among the
graphs, that
generates new graphs from the given two graphs.\\
 Hence, in this
article,  the following section discusses the cutting rules, which
is the basic component for the proposed graph splicing system.
Section 3 deals with the graph splicing system with illustrations.
The section 4 studies some graph theoretical properties of this
system.  The last section concludes with the directions for the
future research in this graph splicing system.

\section{\bf Definitions}
We follow the terminologies and the basic notions of graph theory as
in \cite{bondy} and the terminologies of formal language theory as
in \cite{hopcroft}.\\
 For any finite alphabet $\Sigma$, a labeled graph $G$ over $V$ is
a triple $G = (V,E,L)$ where $V$ is the finite set of vertices(or
nodes), E is finite set of edges of the form $(n,m)$, $n,m \in V, n
\neq m$, where each edge is an unordered pair of vertices and $L$ is
a function from $V$ to $\Sigma$.  An edge $(n,m)$ means that one
end-point of the edge is the vertex $n$ and the other end-point is
the vertex  $m$.  Edge set of $G$ is written as $E(G)$ and the
vertex set of G by $V(G)$.  The number of vertices of a graph is
called the order of the graph and the number of edges of the graph
is called the size of the graph.  We consider only simple graphs
where repeated edges (multiple edges) with same end-points and edges
with both end-points same (loops) are not allowed. The graph $G =
(V,E,\phi)$ refers to an unlabeled graph. We denote an  unlabeled
graph just as $ (V,E)$, instead of $(V, E, \phi)$. Whenever a graph
is considered, we mean only a simple unlabeled graph. We mention
accordingly, when we consider the graphs other than the above one.
\begin{defin} A graph $G$ is said to be in Pseudo-Linear
Form (PLF) if the ordered vertices are positioned  as per the order,
as if they lie along a line  and the edges of the graph drawn
accordingly.
\end{defin}
Ordering of the vertices can be done in any way.  For a particular
ordering, the adjacency matrix of $G$ and the adjacency matrix of
$G$ in PLF, remain the same.  In a graph, the vertices could be
positioned at any place and the edges of the graph drawn
accordingly.  For  the graph in PLF, vertices are first ordered and
positioned as if they lie on a line.  This line may be a horizontal
line or a vertical line or any inclined line.  In case, the line is
horizontal, we can position the ordered vertices either from left to
right or from right to left. So, with out loosing any generality, we
position the vertices from left to right as if the vertices lie on a
horizontal line. Once a graph is in PLF, we name the vertices with a
positive integer that represent their order in the ordering. If a
vertex is second in an ordering, we name that vertex as $2$. So, the
vertex set of $G$ in PLF is $\{1,2,3\ldots \mid V \mid \}$.
 Given an ordering of the vertices, any graph can be redrawn in the
PL form. For example, if the vertices of the graph\\
\includegraphics{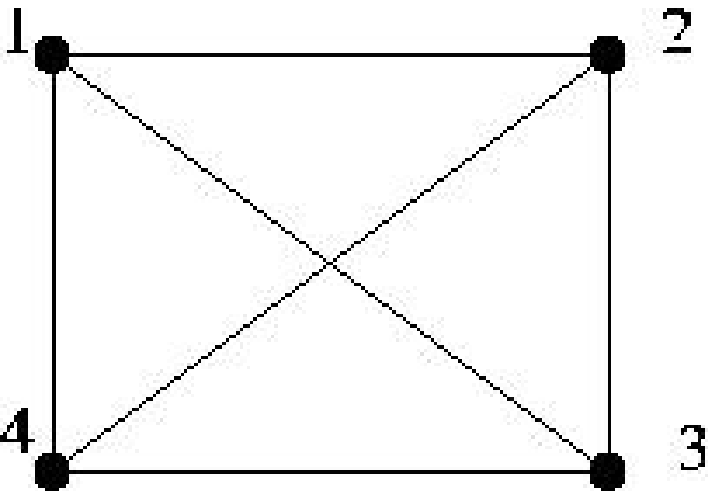}\\
are ordered as $\{1,2,3,4\}$, the corresponding graph in PLF is\\
\includegraphics{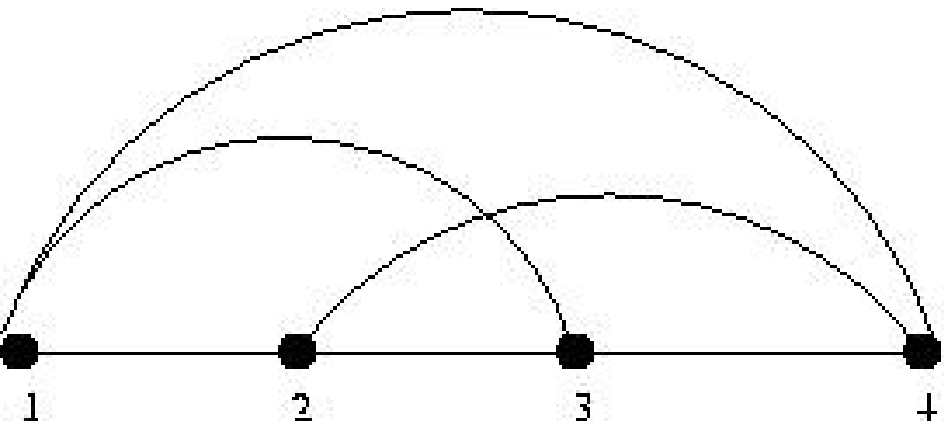}\\
A graph in PLF will look like a path graph with edges going above or
below the linear path. The graph $P_n$ with vertices written
horizontally, is a graph in PLF. From now onwards, unless otherwise
mentioned, we mean a graph as  the one in PLF for some ordering of
the elements of $V$.
\begin{defin}
A cutting rule $\mathcal{C}$ for a graph $G=(V,E)$ in PLF is a pair
$[i,j]$ \footnote{ for the cutting rule $[i,j]$, we use the square
braces and for the edges (i,j), we use the parenthesis}, where $i$
and $j$ are positive integers $ 0<i \leq j \leq |V|, | j - i | \leq
1 $.
\end{defin}
By the condition $ \mid j-i\mid \leq 1 $, we mean that the the
vertices $i$ and $j$ may be  successive vertices (ordered
successively) or both vertices $i$ and $j$ are the same.  A cutting
rule $\mathcal{C}= [i,j]$ is called as a reflexive cutting rule if
$i = j $.
\begin{defin}
The left-degree, $ld_{G}(v)$ of a vertex $v  \in V(G) $ is the
number of edges of $G$ to the left of the vertex $v$ that are
incident with $v$.  The right-degree, $rd_{G}(v)$ of a vertex $v \in
G$ is the number of edges of $G$ to the right of the vertex $v$. The
degree of $v$, $d(v)$,  the number of edges that are incident with
$v$, is the sum of the left-degree and the right-degree of $v$.
\end{defin}
\begin{defin}
Let $V_{l}(v)$ be the set of all vertices that lie to the left of
the vertex $v$ ($ v $ is not in $ V_{l}(v)$).  Similarly,  $
V_{r}(v)$ is the set of all vertices that lie to the right of the
vertex $v$.
\end{defin}
\par{\bf Scheme of cutting}\\

A graph is cut into two parts by cutting some of its edges.The
cutting rule $[i,j]$ cuts a graph $G$ between the vertex $i$ and the
vertex $j$ (if for some reasons, the vertices are named with symbols
other than the positive integers, the cutting rule cuts between the
vertex that comes in the $i^{th}$ position in the ordering and with
the vertex in the $j^{th}$ position). The work of the cutting rule
$[i,j]$ over $G$ is to cut the edge $(i,j)$ and the edges that go
above as well as below the edge $(i,j)$. i.e., The cutting rule
$[i,j]$ cuts the following edges (if they exist in the graph $G$).
\begin{enumerate}
\item The edge $(i,j)$
\item the edges $(i,v), v \in V_{r}(j)$
\item The edges $(v,j), v \in V_{l}(j)$
\item The edges $(u,v), u \in V_{l}(i), v \in V_{r}(j)$
\end{enumerate}
The reflexive cutting rule $(i,i)$  cuts the vertex $i$ and all the
edges that go above as well as  below  the vertex $i$.  i.e., the
reflexive cutting rule $i$ cuts the following.
\begin{enumerate}
\item The vertex $i$
\item The edges $(u,v), u \in V_{l}(i), v \in V_{r}(j)$
\end{enumerate}
When an edge $(i,j)$ is cut into two parts, we call the the two
parts of the edge as hanging-edges or free-edges.  Similarly, when a
vertex is cut, we call that vertex as a hanging-vertex or a
free-vertex. If an edge $(i,j)$ is cut, we write the left part of
the edge as $(i,j]$ (indicating that the free-end is the right end)
and the right part of the edge as $[i,j)$ (indicating that the
free-end is the left end).  The edges $(i,j]$ and $[i,j)$ are drawn
as illustrated with a $\times$ at their free ends. If a vertex $v$
is cut, the left part of the vertex is written as $v]$ and the right
part is written as $[v$. $[v]$ indicates just that
the vertex $v$ is cut.\\
\begin{center}
\includegraphics{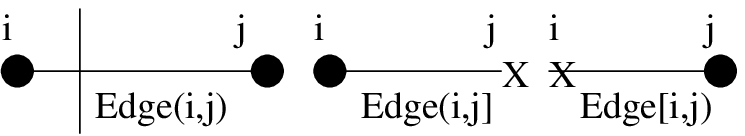}
\end{center}

The set $ECUT_{G}(\mathcal{C})$ represents the set of all edges of
$G$ that got cut by the cutting rule $\mathcal{C}$ and the $\mid
ECUT_{G}(\mathcal{C}) \mid$ (the cardinality of the set) is the
power of the cutting rule $ \mathcal{C}$ with respect to the graph
$G$. Power of a cutting rule with respect to $G$ indicates the
number of edges that got cut by that cutting rule in $G$.  The set
$VCUT_{G}(\mathcal{C})$ represents the set of all vertices that got
cut by the vertex $v$. Only for the reflexive cutting rules, the set
$VCUT_{G}(\mathcal{C})$ will exist and for all the other cutting
rules, this set is $\phi$.  Since any reflexive cutting rule can cut
only one vertex, the set $VCUT_{G}(\mathcal{C})$ is always
singleton. For a reflexive cutting rule, the set
$ECUT_{G}(\mathcal{C})$ can be $\phi$ ( means that no edge is going
above or below the vertex $i$ in the graph $G$ ).
\\
When a graph $G$ is cut into two by a cutting rule $\mathcal{C}$,
the left part of the  graph is called as $Prefix(G)$ and the right
part is called as $Suffix(G)$.  Obviously, $ECUT_{G}([i,j]) =
ECUT_{G}([i,i] \cup ECUT_{G}([j,j] \cup \{(i,j)\}.$

We illustrate the cutting of the graph $K_5$, a complete graph with
five vertices using the cutting rule $[2,3]$.\\
\includegraphics{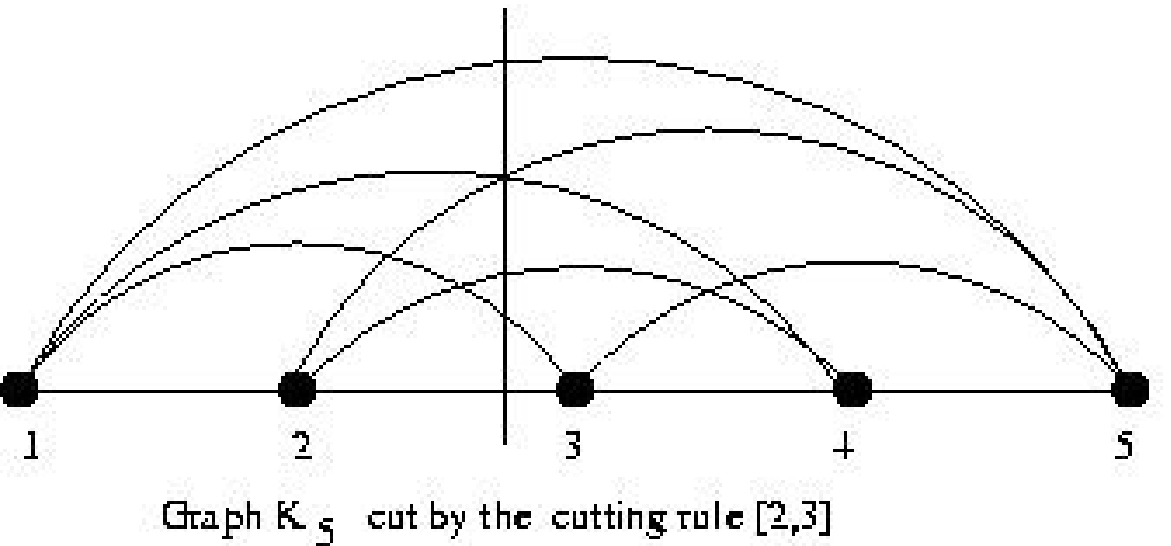}\\
\includegraphics{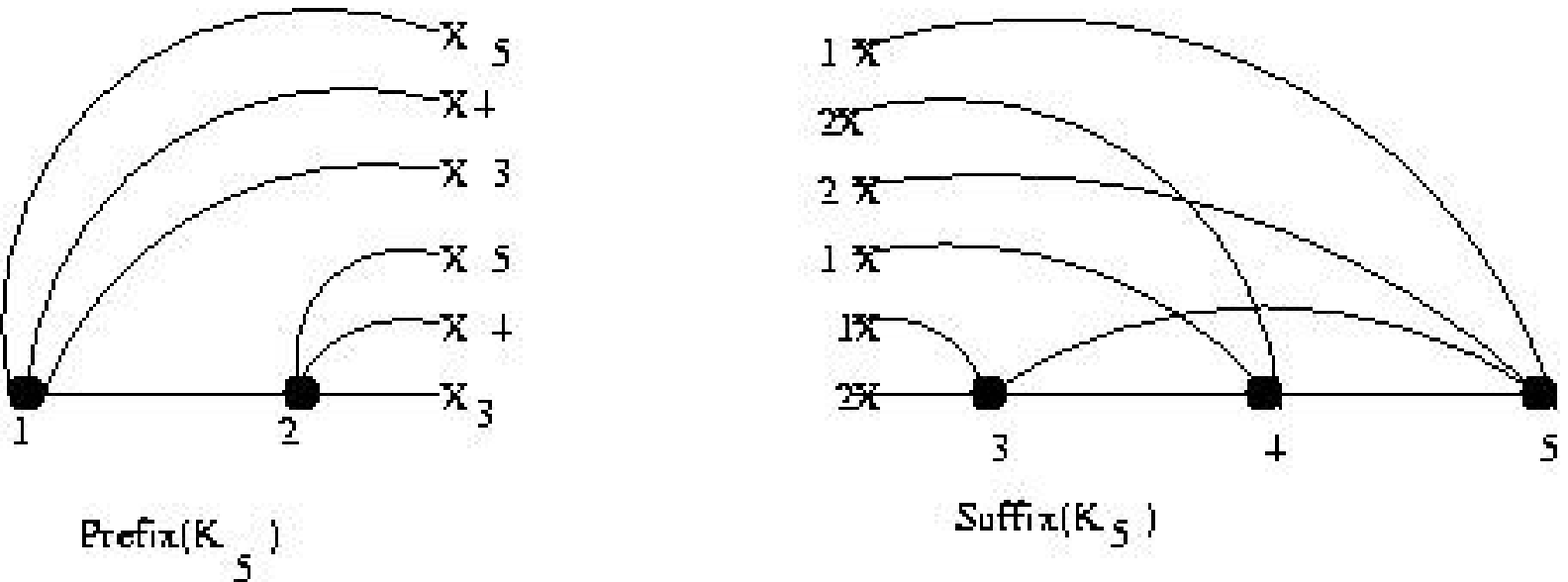}\\
The $prefix(K_5)$ and $Suffix(K_5)$ are also graphs with the vertex
set \[V(Prefix(K_5) = \{1,2\}\] and with the edge set
 \[E(Prefix(K_5)=
\{ (1,2), (1,3],(1,4],(1,5],(2,3],(2,4],(2,5]\}.\]
  Similarly,
$V(Suffix(K_5) = \{3,4,5\}$ and $E(Suffix(K_5)= \{ [2,3),[2,4),
[2,5),$ $ [1,3),[1,4),[1,5), (3,4),(3,5)\}$. $ECUT_{K_5}([2,3]) =
\{(1,3),(1,4),(1,5),(2,3),$\\ $(2,4),(2,5)\}$. $VCUT_{K_5}([2,3])=\phi$.

\section{Graph Splicing System}
\begin{defin}
A splicing rule $\mathcal{S} = (\mathcal{C}_1,\mathcal{C}_2)$, is a
pair of cutting rules.
\end{defin}

Given  two graphs $G$,$H$ and a splicing rule $\mathcal{S} =
(\mathcal{C}_1,\mathcal{C}_2)$, the first graph $G$ is cut as
specified by $\mathcal{C}_1$ and the second graph $H$ is cut as
specified by $\mathcal{C}_2$.  As a result we get the four
cut-graphs viz., $Prefix(G)$,$Suffix(G)$,$Prefix(H)$ and
$Suffix(H)$.
\\\\
{\bf Mode of recombination}\\
\begin{defin}
$Prefix(G)$ (or $Prefix(H)$) recombines with the $Suffix(H)$ (or
$Suffix(G)$) if and only if $\mid ECUT_{G}(\mathcal{C}_1) \mid$
=$\mid ECUT_{H}(\mathcal{C}_2) \mid$ and $\mid
VCUT_{G}(\mathcal{C}_1) \mid$ =$\mid VCUT_{H}(\mathcal{C}_2) \mid$.
In other words, for a recombination, the number of hanging-edges in
$Prefix(G)$ (or $Prefix(H)$) should be the same as that of the
number of hanging-edges in $Suffix(H)$(or $Suffix(G)$) and the the
number of hanging-vertices in $Prefix(G)$ (or $Prefix(H)$) should be
the same as that of the number of hanging-vertices in $Suffix(H)$(or
$Suffix(G)$).
\end{defin}
The above definition tells that for a splicing process to end up in
a recombination, the power of both the cutting rules present in the
splicing rule $\mathcal{S}$ should be the same. We assign a positive
integer, called the power of the splicing rule,to every splicing
rule $\mathcal{S} = (\mathcal{C}_1,\mathcal{C}_2) $ if and only if
the powers of $\mathcal{C}_1$ and $\mathcal{C}_2$ are the same and
the common value is the power of the splicing rule  $\mathcal{S}$.
Further, if one cutting rule in $\mathcal{S}$ is reflexive, the
other should also be reflexive.
\begin{defin}
 Every  hanging-edge of the $Prefix(G)$ (or $Prefix(H)$) recombines (or joins) with only one
 hanging-edge of the $Suffix(H)$ (or $Suffix(G)$), and every hanging-edge of
 $Suffix(H)$ (or $Suffix(G)$) has the recombination
  with only one hanging-edge of $Prefix(G)$ (or $Prefix(H)$). The
    hanging-vertex (if available) of the $Prefix(G)$ (or $Prefix(H)$) recombines with
    the hanging-vertex of $Suffix(H)$(or $Suffix(G)$).
 \end{defin}
Thus, $Prefix(G)$ recombines with $Suffix(H)$ to generate new
graphs. After the recombination, we order the vertices  of the new
graph (this will be in PL form) in the same sequence as it appears
and name them accordingly. New graphs are generated because of the
recombination of the edges that are cut. If there are more than one
hanging-edges in both $Prefix(G)$ and  $Suffix(H)$, the
hanging-edges of the $Prefix(G)$ can recombine with the
hanging-edges of $Suffix(H)$ in more than one way. If there are $m$
hanging-edges in both $Prefix(G)$ and $Suffix(H)$,  the
hanging-edges can recombine in $m!$ ways, generating $m!$ new
graphs. In other words, the number of such recombinations is just
the number of bijective mappings from the set
$ECUT_{G}(\mathcal{C}_1)$ to the set $ECUT_{H}(\mathcal{C}_2)$. When
the $Prefix(H)$ recombines with the $Suffix(G)$, the same number of
$m!$ will be generated.  Thus, the splicing  of two graphs $G$ and
$H$ using a splicing rule of order $m$, generates $2(m!)$ new
graphs.
\\Thus, splicing process comprises of cutting as well as the
recombination. If the splicing of $G$ and $H$ using $\mathcal{S}$
generates a new graph $F$ by the recombination of the $Prefix(G)$
with the $Suffix(H)$, we denote that by $G{\vdash^1_{\mathcal{S}}} H
= F$ (indicating that $F$ is the first splicing product). Similarly,
$G{\vdash^2_{\mathcal{S}}} H = F$ indicates that $F$ is generated by
the recombination of the $Prefix(H)$ with the $Suffix(G)$
(indicating that this F is the second product of splicing).  Just
$G{\vdash_{\mathcal{S}}} H = F$ indicates that $F$ may be either the
first splicing product or the second splicing product. The splicing
scheme(process) is denoted by $\sigma$. For a splicing process, one
requires two graphs and a splicing rule. The set of all graphs
generated by splicing $G$ and $H$ using the splicing rule
$\mathcal{S}$ is denoted by $\sigma(\{G,H\},\mathcal{S})$.
Similarly, $\sigma_1(\{G,H\},\mathcal{S})$,
$\sigma_2(\{G,H\},\mathcal{S})$ are meant  accordingly.
\begin{defin}
 The Graph Splicing System $ \nu = (\mathfrak{A},\mathfrak{S})$, where
\begin{enumerate}
\item[$\mathfrak{A}$] A finite set of simple, unlabeled graphs, called the set of axioms.

\item [$\mathfrak{S}$] A finite set of splicing rules.

\end{enumerate}
The underlying splicing scheme is $ \sigma(\{G,H\},\mathcal{S})$, $
G ,H \in \mathfrak{A}, \mathcal{S} \in \mathfrak{S} $.
\\
The set of all graphs generated by splicing all pairs of the graphs
of $\mathfrak{A}$ with all splicing rules of $\mathfrak{S}$ (The
graph language of the splicing system $\nu$),
\begin{equation*}
 L(\nu) =  \sigma(\mathfrak{A}) = \bigcup _{G,H \in \mathfrak{A},\\\mathcal{S}\in
\mathfrak{S}} \sigma(\{G,H\},\mathcal{S})
\end{equation*}
\end{defin}

In the DNA recombination, when some restriction enzymes and a ligase
are present in a test tube, they do not stop after one cut and paste
operation, but they act iteratively. The products of a splicing
again take part in the splicing process.  For an iterative splicing
among the graphs, the axiom set should contain many copies of the
same element.  Ordinary sets are composed of pairwise different
elements, i.e., no two elements are the same.  If we relax this
condition, i.e., if we allow multiple but finite occurrences of any
element, we get a generalization of the notion of a set which is
called a {\it multiset}. We assume that our axiom set is a multiset.
That means infinitely many copies of the elements of the axiom set
will be present in the set, which facilitates the elements to take
part in the splicing process iteratively.  Even the product of a
splicing  process will also be available infinite number of times.
To make a graph splicing system into an iterated graph splicing
system, the only requirement is to make the axiom set $\mathfrak{A}$
into a multiset such that infinitely many copies of the elements of
$\mathfrak{A}$ are in     $\mathfrak{A}$.

\begin{defin}
The graph language of an iterative graph splicing system $ \nu =
(\mathfrak{A},\mathfrak{S})$, where $\mathfrak{A}$ is a multiset
such that infinitely many copies of the elements of $\mathfrak{A}$
are in $\mathfrak{A}$, is defined as $L(\nu)=
\sigma^*(\mathfrak{A})$ where

\begin{eqnarray*}
\sigma^0(\mathfrak{A}) &=& \mathfrak{A},\\
\sigma^{i+1}(\mathfrak{A})&=&\sigma^i(\mathfrak{A}) \cup \sigma(\sigma^i(\mathfrak{A}),\\
\sigma^*(\mathfrak{A})&=&\bigcup_{i\geqslant0}
\sigma^i(\mathfrak{A}).
\end{eqnarray*}
\end{defin}

\begin{eg}
Consider the   graph splicing system $\nu = (\{C_3,C_4\}   ,\{ (
[1,2],[2,3])\})$. $C_3$ and $C_4$ are the cycle graphs of order $3$
and $4$ respectively.
\[
L(\nu)= \sigma (  \{C_3,C_4\},\mathcal{S}) \cup \sigma (
\{C_4,C_3\},\mathcal{S}) \cup \sigma (  \{C_3,C_3\},\mathcal{S})
\cup \sigma (  \{C_4,C_4\},\mathcal{S})\] where  $\mathcal{S}$ is
the splicing rule $( [1,2],[2,3])$.  The power of the splicing rule
is 2.  In each splicing process, $2(2!)=4$ new graphs will be
generated.  So, $L(\nu)$ will have a total of 16 new graphs. Of
these, some of the graphs are isomorphic to each other. It is found
that the  non-isomorphic graphs in $L(\nu)$ are $C_3, C_4, C_5$ and
a graph $G$, where $G = ( \{1,2\}, \{(1,2),(1,2)\})$. The above
graph $G$ is not a simple graph ( it has a multiple edge between the
vertices $1$ and $2$). This makes us to conclude that the splicing
of two simple graphs need not be simple.
\end{eg}
\section{Properties}
\begin{prop}
Given a graph $G$, the power of the cutting rule $[i,j]$ with
respect to the graph $G$ is
\begin{equation*}
  rd(i)-ld(i) + \sum_{v \in V_l(i)} (rd(v)-ld(v))  .
\end{equation*}
\end{prop}

\par

Proof: Let $G$ be the given graph. We count the total  the number of
edges in $G$ that got cut by the cutting rule, which is the power of
the cutting rule.  We classify the proof into two cases based on the
existence of the edge $(i,j)$ in $G$ or not.
\\
{\bf Case(i) : $(i,j) \in E(G)$}.  \\
We know that the cutting rule [i,j] cuts the following edges.

 \begin{enumerate}
 \item The edge $(i,j)$
 \item The edges $(i,v), v \in V_{r}(j)$
 \item The edges $(v,j), v \in V_{l}(j)$
 \item The edges $(u,v), u \in V_{l}(i), v \in V_{r}(j)$
 \end{enumerate}
The expression
\begin{equation}
rd(i)+ld(j) - 1
\end{equation}
brings out the number of edges which fall under (1),(2) and (3) in
the list above.  Since the edge  (i,j) is counted in both $rd(i)$ as
well as in $ld(j)$, we subtract  one from the
expression.\\
Let $A$ be the set of edges whose left end is $V_l(i)$. Let $B
\subset A$, be the set of edges of $A$ whose right end is in
$V_l(i)$. i.e.,  both the ends of edges  in $B$ are in    $V_l(i)$.
Let $C \subset A$ be the set of edges of $A$ whose right  end is $i$
i.e., for the edges in $C$, one end is $V_l(i)$and the other end is
$i$. Let $D \subset A$, be the set of edges of $A$ whose right end
is $j$. i.e., for the edges in $D$ one end is in $V_l(i)$ and the
other end is $j$. Let $E$ be the set of edges of $A$ whose right
 end is
in $V_r(j)$ i.e., the set of edges whose left end is in $V_l(i)$ and
the right end is in $V_r(j)$ Obviously, the set of edges which come
in (4) in the above list, will be $E$.
 \begin{equation*}
\mid A \mid = \sum_{v \in V_l(i)} rd(v); \mid B \mid = \sum_{v \in
V_l(i)} ld(v); \mid C\mid = ld(i ); \mid  D \mid = ld(j)-1.
\end{equation*}
 Since the edge $(i,j)$ would be counted in $ld(j)$, we subtract
one from $ld(j)$.

Number of edges that come under (4) is
\begin{equation}
 \mid E \mid =  \mid A \mid  - \mid B \mid  - \mid C \mid  - \mid D \mid   \\
= \sum_{v \in V_l(i)} rd(v) -   \sum_{v \in V_l(i)} ld(v) -ld(i) -
(ld(j)-1)
\end{equation}

Hence, the total number of edges cut by $[i,j]$
\begin{equation*}
= (1) +  (2) = rd(i)-ld(i) + \sum_{v \in V_l(i)} (rd(v) - ld(v))
\end{equation*}
{\bf Case(ii) : $(i,j)~  not ~in~  E(G)$}\\
We proceed similarly as the case(i). The number of edges that come
under (1),(2) and (3) is
\[rd(i) + ld(j)\]
  Number of edges that come under (4) is
  \begin{equation*}
   = \sum_{v \in V_l(i)} rd(v) -   \sum_{v \in V_l(i)} ld(v) -ld(i) - ld(j)
 \end{equation*}
 Hence, the total number of edges cut by $[i,j]$
 \begin{equation*}
 =  rd(i)-ld(i) + \sum_{v \in V_l(i)} (rd(v) - ld(v))
  \end{equation*}
  In both the cases, we get the same expression.  Hence the proof.

\begin{theorem}

   In any graph $G$, the sum of the differences between the right degree and the left degree
  of all the vertices is zero.
  \end{theorem}

  {\bf Proof:} In the Proposition 1, in computing the power of a cutting rule
  $[i,j]$,
  we counted the number of edges whose one end is in $V_l(i)$ and the other end is in $V_r(j)$
  by deleting some edges from the set $A$ which is the set of edges whose left end is in
  $V_l(i)$. Instead, we can have the set $A$ to be the set of edges whose right end is in
  $V_r(j)$ and proceed in an analogous way, as in the proof of Proposition 1. We get the
  power of the cutting rule $[i,j]$ to be

  \begin{equation*}
   ld(j)-rd(j) + \sum _{v \in V_r(j)} ld(v) - rd(v)
   \end{equation*}
   which is a symmetric one with the expression got in Proposition 1.\\
   Since the power of a cutting rule is a constant with respect to a $G$,
   both the expressions should be equal.\\
   \[ rd(i)-ld(i) + \sum_{v \in V_l(i)} (rd(v) - ld(v)) = ld(j)-rd(j) +
    \sum_{v \in V_r(j)} (ld(v) - rd(v))\]  \\
     \[implies ~~ \sum_{v \in V} (rd(v) - ld(v))= 0~~ or ~~ \sum_{v \in V} (rd(v) - ld(v))= 0
    \]

    \begin{corollary}

    The number of edges in a graph $G$ is always  \[\sum_{v \in V}rd(v)\] or \[\sum_{v \in V}
    ld(v)\]

      \end{corollary}
      \par
      {\bf Proof}
      \[
      \sum d(v) = \sum (ld(v) + rd(v)) = 2 \mid E \mid \]   \\

      we have       \[ \sum (ld(v) - rd(v)) = 0.\]            \\
      This  implies, \[ \sum ld(v) = \mid E \mid =  \sum rd(v)\].

      \begin{rem}
      The above Corollary can also be proved in another way using the fact that
      every edge should contribute one to the left degree of some vertex and one to the
      right degree of some other vertex.
      \end{rem}
          For want of space,We state some of the results without proofs.

       \begin{theorem}
       \begin{enumerate}
       \item $G\vdash H  =  H  \vdash_{S^R} G$ ,   where $S^R$ is the splicing rule in which the cutting rules of S
       got swapped.
       \item $G \vdash _S H  \neq   H \vdash _S G$ i.e.,the  splicing operation is not
       commutative.
       \item   The splicing operation preserves the degrees of the vertices
       \item    Regularity is preserved by the splicing.  i.e., if we splice any two
       regular graphs, the splicing product is again a regular graph.
       \item maximum size of the splicing product of G and H will be the sum of the
       orders of $G$ and $H$ minus 1.
       \item For a complete graph $K_n$, $rd(i) = ld(n+1-i)$, for every
       vertex $ i \in V(K_n)$.
       \item The set of all simple graphs is not closed with respect
       to the splicing operation.
       \end{enumerate}
       \end{theorem}

       \begin{theorem}
       A graph $G$ is said to contain a cycle if and only if there exists a sequence $A$ of
       successive cutting rules \footnote{The rules $[i,i+1]$ and $[i+1,i+2]$  are termed
        successive cutting rules} with power $> 1$ such that \[\bigcap_{[i,i+1]\in A}ECUT_G
         ([i,i+1]) \neq \phi\].

        \end{theorem}
        \begin{theorem}
         Let $ G $ and $H$ be any two isomorphic graphs.  Let $G \vdash_{\mathcal{S}} H$ =
         $F$, for any splicing rule $\mathcal{S}$.
         Then F is isomorphic to
        $G$ (or H) if and only if the order of the graph $F$ and the order of $G$ (or the order
        of $H$) are the same.
        \end{theorem}
        \begin{theorem}
         If for a  graph $G$, there exists only one  cutting rule
        whose power is equal to the size of the graph $G$, Then $G$
        is bipartite.
        \end{theorem}

        \section{conclusion}
        As graphs are better suited for representing complex structures, a model for splicing
        the graphs, graph splicing system is introduced, which can be applied to all types of graphs.
        Though the graph splicing is introduced as a new operation among the graphs, studying the computational
        effectiveness of this graph splicing system is an important area to explore.  One can introduce various parameters like
        the number of graphs in the axiom, the number of splicing rules, power of the splicing rule etc.,
        and finding the minimum value of the parameters for which the graph splicing system is still
        computationally complete.  Besides, as a new line of thinking, a nice investigation to bring out
        the utility of the splicing in graph theory is worth.

 \end{document}